\documentclass[11pt,a4paper]{article}
\usepackage{tikz}
\usetikzlibrary{decorations.pathmorphing,positioning,decorations.pathreplacing,shapes,calc}
\tikzset{snake it/.style={decorate, decoration={snake,amplitude = .5mm,segment length=1.95mm}}}
\tikzset{zigzag it/.style={decorate, decoration=zigzag}}
\def\centerarc[#1](#2)(#3:#4:#5)
    { \draw[#1] ($(#2)+({#5*cos(#3)},{#5*sin(#3)})$) arc (#3:#4:#5); }
\tikzset{gluon/.style={decorate, draw=black,
    decoration={coil,amplitude=4pt, segment length=5pt}}}
\usepackage{jheppub}
\usepackage{subfigure}
\usepackage{xspace}
\usepackage{color}

\usepackage{soul,xcolor}
\setstcolor{red}

\definecolor{math1}{rgb}{0.368417,0.506779,0.709798}
\definecolor{math2}{rgb}{0.880722,0.611041,0.142051}
\definecolor{math3}{rgb}{0.560181,0.691569,0.194885}
\definecolor{math4}{rgb}{0.922526,0.385626,0.209179}
\definecolor{math5}{rgb}{0.528488,0.470624,0.701351}
\definecolor{navyblue}{rgb}{.18,.18,.56}
\newif\ifstartedinmathmode
\DeclareRobustCommand{\gotomathmode}[1]{\relax\ifmmode\startedinmathmodetrue\else\startedinmathmodefalse\fi
\ifstartedinmathmode #1 \else $#1$\fi}

\makeatletter
\newcommand{\vast}{\bBigg@{4}}
\newcommand{\Vast}{\bBigg@{5}}
\makeatother

\newcommand{\D}{\mathrm{d}}
\newcommand{\I}{i}

\newcommand{\MS}{\gotomathmode{\overline{\rm MS}}\xspace}

\newcommand{\eps}{\gotomathmode{\epsilon}\xspace}
\newcommand{\Neps}{\gotomathmode{n_\epsilon}\xspace}

\newcommand{\neqcd}{\gotomathmode{\Neps{\rm QCD}}\xspace}

\def\CDR{\text{\scshape cdr}\xspace}
\def\HV{\text{\scshape hv}\xspace}
\def\FDH{\text{\scshape fdh}\xspace}
\def\DRED{\text{\scshape dred}\xspace}
\def\DREG{\text{\scshape dreg}\xspace}
\def\io{{\rm i0}^+}

\newcommand{\Zjet}{\gotomathmode{Z_{q}}\xspace}
\newcommand{\Zantijet}{\gotomathmode{\bar{Z}_{q}}\xspace}
\newcommand{\soft}{\gotomathmode{\mathcal{S}}\xspace}
\newcommand{\softHQ}{\gotomathmode{\mathcal{S}'}\xspace}
\newcommand{\FMms}{\gotomathmode{F_1(M,m,s)}\xspace}
\newcommand{\FMzs}{\gotomathmode{F_1(M,0,s)}\xspace}
\newcommand{\Fmms}{\gotomathmode{F_1(m,m,s)}\xspace}
\newcommand{\Fzzs}{\gotomathmode{F_1(0,0,s)}\xspace}
\newcommand{\order}[1]{\gotomathmode{\mathcal{O}\left(#1\right)}}

\newcommand{\ie}{i.\,e.\ }

\begin{document}
\thispagestyle{empty}
\begin{flushright}
PSI-PR-18-13\\
ZU-TH 40/18\\
\today\\
\end{flushright}
\vspace{3em}
\begin{center}
{\Large\bf
Small-mass effects in heavy-to-light form factors
}
\\
\vspace{3em}
{\sc T. Engel$^{a}$, C.~Gnendiger$^{a}$, A.~Signer$^{a,b}$, Y. Ulrich$^{a,b}$
}\\[2em]
{\sl ${}^a$ Paul Scherrer Institut,\\
CH-5232 Villigen PSI, Switzerland \\
\vspace{0.3cm}
${}^b$ Physik-Institut, Universit\"at Z\"urich, \\
Winterthurerstrasse 190,
CH-8057 Z\"urich, Switzerland}
\setcounter{footnote}{0}
\end{center}
\vspace{2ex}
\begin{center}
\begin{minipage}[]{0.9\textwidth} {} 
{\sc Abstract:} We present the heavy-to-light form factors with two
different non-vanishing masses at next-to-next-to-leading order and
study its expansion in the small mass. The leading term of this
small-mass expansion leads to a factorized expression for the form
factor. The presence of a second mass results in a new feature, in
that the soft contribution develops a factorization anomaly. This
cancels with the corresponding anomaly in the collinear contribution.
With the generalized factorization presented here, it is possible to
obtain the leading small-mass terms for processes with large masses,
such as muon-electron scattering, from the corresponding massless
amplitude and the soft contribution.
\end{minipage}
\end{center}

\setcounter{page}{1}

\newpage

\section{Introduction}
\label{sec:introduction}

Perturbative calculations in QED and QCD are often carried out setting
fermion masses to zero. This reduces the number of scales and results
in a simplification of the computation of the virtual corrections.
However, the neglect of fermion masses also has a profound impact on
the structure of the infrared (IR) singularities. In the case of QED,
fermion masses $m\neq 0$ can be considered as a regulator for
collinear singularities and only soft (and ultraviolet) singularities
remain. Collinear singularities, that in dimensional regularization
are manifest as poles $(1/\eps)^n$, are then replaced by terms
$\log(m^2/s)$, where $s$ is a kinematic invariant. Due to the
non-Abelian nature of gluon interactions the situation is more
involved in QCD. However, the structure of IR singularities in
massless theories~\cite{Becher:2009qa, Gardi:2009qi} is altered
through quark masses~\cite{Becher:2009kw} in a similar way.

It is not possible to obtain a complete amplitude with $m\neq 0$ from
the corresponding amplitude with $m\!=\!0$ without performing a full
computation. The universal structure of IR singularities, however,
enables the extraction of the leading $\log(m^2/s)$ terms from the
massless amplitudes. If the mass is small compared to all other
kinematic invariants, this can provide a reasonable approximation. In
fact, such relations have been worked out already at
next-to-next-to-leading order (NNLO). Initially this was done for QED
in the context of Bhabha scattering~\cite{Penin:2005eh}. Later, a more
general approach has been presented~\cite{Mitov:2006xs,Becher:2007cu}
that relies on factorization and is also valid in QCD. Very recently,
these considerations have been extended beyond NNLO, in particular for
the heavy-quark form factor~\cite{Liu:2017axv, Liu:2018czl,
Blumlein:2018tmz}.

In this article we extend the previous NNLO analyses to the case of
two different external masses with $M\!\gg\!m$. We provide an IR
factorization formula that allows to obtain the logarithmic
corrections $\log(m^2/M^2)$ and $\log(m^2/s)$ from the amplitudes with
$m\!=\!0$.  As it turns out, diagrams with fermion loops play a
special role and lead to complications that are absent if there is no
second (large) mass.

This extension is particularly relevant in light of the recently
proposed experiment {\sc MUonE} at CERN to measure muon-electron
scattering at high precision.  Precise data of the angular
distribution of this process will provide an alternative approach to
measure the hadronic vacuum
polarization~\cite{Abbiendi:2016xup,Abbiendi:2017xap}. In order to
achieve this, the theoretical predictions have to be known with
sufficient accuracy. While a fully differential NLO calculation is
available~\cite{Alacevich:2018vez} and the contribution of the
hadronic vacuum polarization has been evaluated at
NNLO~\cite{Fael:2018dmz, Fael:2019nsf}, a full NNLO computation is
still missing.  So far, the two-loop integrals of $\mu\!-\!e$
scattering have been calculated with vanishing electron
mass~\cite{Mastrolia:2017pfy,DiVita:2018nnh}.  This enables the
computation of the two-loop amplitude for massless electrons. However,
the necessary theoretical accuracy of $10^{-5}$ requires the inclusion
of the leading electron mass effects.

As a concrete example, we consider the effects of a small final-state
mass to the heavy-to-light form factor, \ie the transition between a
heavy fermion of mass $M$ to a lighter one of mass $m$.  While one
motivation is to obtain the generalized factorization formula
mentioned above, the heavy-to-light form factor is an interesting
quantity in itself. It is an important ingredient in several precise
probes of the Standard Model. In the context of the muon decay, it is
one of the most precisely measured quantities in particle physics and
is used to extract for example the Fermi constant $G_F$. But also $b$
and top quark decays play an important role in precision tests of the
Standard Model.

Therefore, the QED and QCD corrections to the heavy-to-light form
factor are needed at high precision. At the one-loop level, the QED
correction to the muon decay with full final-state mass dependence has
been known for many decades~\cite{Kinoshita:1958ru,PhysRev.101.866}.
The muon decay branching ratio was then calculated at two-loops in QED
using the optical theorem~\cite{vanRitbergen:1999fi}. The electron
energy spectrum including leading electron mass effects was obtained a
few years later~\cite{Arbuzov:2002pp,Arbuzov:2002cn}.  In 2008 the
full electron mass dependence of the energy spectrum was computed
numerically at NNLO in QED~\cite{Anastasiou:2005pn}. For vanishing
final-state masses, analytic expressions for two-loop QCD corrections
to the heavy-to-light form factors have been
calculated~\cite{Bonciani:2008wf,Asatrian:2008uk,Beneke:2008ei,Bell:2006tz}
in several regularization schemes~\cite{Gnendiger:2016cpg}.

Recently, the master integrals needed to compute the form factors for
arbitrary masses have been presented~\cite{Chen:2018dpt} in terms of
generalized polylogarithms (GPL)~\cite{Goncharov:1998kja}.  Using
these integrals, we calculate the NNLO form factor for two different
non-vanishing fermion masses. However, the numerical evaluation of the
GPLs is time consuming and, therefore, difficult to implement in a
full NNLO Monte Carlo. Fortunately, the physical mass ratios
$m_e/m_\mu$, $m_b/m_t$, and $m_s/m_b$ are small enough to allow for an
expansion in the small fermion mass. We perform such an expansion,
write it in a factorized form and compare it to the result with the
full mass dependence. All calculations are conducted in a generic way
that allows to extract the results in various dimensional
regularization schemes.
 
This paper is structured as follows: In Section~\ref{sec:notation} we
introduce our notation and describe our setup, including a review of
the regularization schemes used. Then, in Section~\ref{sec:resultstop}
we discuss our results for the mass effects in the heavy-to-light form
factors and present the factorization. Lengthy results are provided in
an ancillary file. In Section~\ref{sec:resultshq} we compare our
results to previous factorization formulations and discuss the
extension of our approach to other processes like the heavy-quark form
factor and $\mu\!-\!e$ scattering. Finally, we conclude in
Section~\ref{sec:conclusion}.

\section{Notation and setup}
\label{sec:notation}
The decay of a fermion with momentum $p^2\!=\!M^2$ into a lighter
fermion with momentum $q^2\!=\!m^2$ is closely related to the
amplitude $\mathcal{A}^\rho = \bar u(q) \,\Gamma^\rho\,u(p)$. The
vertex can be written in terms of form factors $F_i$ and $G_i$ as
\begin{align}
\Gamma^\rho &= 
  F_1\, \gamma^\rho
+ \frac{\I}{2M}\, \sigma^{\rho\sigma}\, Q_\sigma\, F_2
+ \frac{1}{2M}\, Q^\rho F_3\nonumber\\&\quad
+ G_1\, \gamma_5\gamma^\rho
+ \frac{\I}{2M}\, \gamma_5\,\sigma^{\rho\sigma}\, Q_\sigma\, G_2
+ \frac{1}{2M}\, \gamma_5\, Q^\rho G_3\,,
\label{eq:ff}
\end{align}
with $Q=p-q$. There are different conventions used in the literature
for the form factors $G_i$. In \eqref{eq:ff} they are defined such
that each $G_i$ is closely related to the corresponding $F_i$.
Exploiting the symmetry of the Lagrangian under the simultaneous
transformation of the light fermion field $\psi_q\to\gamma_5\psi_q$
and $m\to-m$, the axial form factors can be obtained from the vector
form factors through~\cite{BERMAN196220}
\begin{align}
G_i(m) = F_i(-m)\,.\label{eq:bermansym}
\end{align}
Therefore, we restrict ourselves to the discussion of the $F_i$.
Furthermore, since it has the most interesting IR structure, we focus
on $F_1$ and comment on $F_2$ and $F_3$ occasionally.

We parametrize the result in terms of the three kinematic invariants
of the process
\begin{align}
M\,,
\quad
z=\frac mM\,,
\quad\text{and}\quad
x=\frac{Q^2}{M^2}\,.
\end{align}
For convenience we also define
\begin{align}
  \label{eq:sdef}
  s = 2\,p\cdot q = M^2(1-x) + m^2
\end{align}
and write the dependence on the kinematics of the form factors as
$F_i(M,m,s)$. 

The ultraviolet (UV) renormalization of the masses and wave functions
of the external particles is performed in the on-shell scheme, while
the QED and QCD couplings are renormalized in the \MS scheme. The
renormalization constants
from~\cite{Gnendiger:2016cpg,Gnendiger:2014nxa,Broadhurst:1991fy} are
extended to account for $n_f$ massless flavours, $n_m$ flavours with
mass $m$, and $n_h$ flavours with mass~$M$.

The form factors $F_i$ in~\eqref{eq:ff} receive contributions from one
one-loop and 9 (13) two-loop diagrams in QED (QCD), including fermion
loops with masses $M$ and $m$ as well as massless ones. Writing all
tensor integrals in terms of scalar ones, this results in about 700
scalar loop integrals. Using the program {\tt
reduze}~\cite{vonManteuffel:2012np} which implements Laporta's
algorithm~\cite{Laporta:2001dd}, these reduce to a set of 40 two-loop
and three one-loop master integrals.

\subsection{Expansion in the small mass} 
\label{sec:mexp}

The form factors can be written as a function of the three invariants
$m$, $M$, and $s$. To obtain the full dependence on all invariants, we
use the integrals from~\cite{Chen:2018dpt}. However, as mentioned
above, the numerical evaluation of the
GPLs~\cite{Vollinga:2004sn,Frellesvig:2016ske} is rather time
consuming. For physical applications we therefore consider an
expansion in
\begin{align}
m\ll M \sim s\sim Q^2\,,
\label{eq;scales}
\end{align}
which corresponds to an expansion in $z$.  Obviously, the algebraic
part of the expression can be expanded trivially. In order to obtain
the expanded form of the master integrals, we use the method of
regions~\cite{Beneke:1997zp}.

Introducing two light-like directions $e_\mu=(1,0_\perp,+1)/\sqrt2$
and $\bar e_\mu=(1, 0_\perp,-1)/\sqrt2$ with a $(d-2)$-dimensional
perpendicular component\footnote{Note that this definition differs
from the usual light-cone vector by a factor of $\sqrt2$.}, we
parametrize the external momenta as
\begin{subequations}\label{eq:mompar}
\begin{align}
q^\mu &= \sqrt{2} E\, e^\mu + q^\mu_\perp \,,\\
p^\mu &= \frac{M}{\sqrt{2}}  (e^\mu + \bar e^\mu)\,, 
\end{align}
\end{subequations}
with $(q_\perp)^2\!=\!m^2$ and $s\!=\!2 M E$. To obtain the expansion,
we attach a small scaling parameter $\lambda\ll 1$ to $m$ and
$q_\perp$. In light-cone coordinates $k\!=\!  (k_+,k_-,k_\perp)$ with
$k_+\!=\!(k\cdot e)$ and $ k_-\!=\!(k\cdot \bar e)$ the external
momenta then scale as $q\!\sim\!(0,1,\lambda)$ and $p\!\sim\!(1,1,0)$.
The momentum regions that contribute to the integrals are associated
with the following scalings of the integration momenta: 
\begin{subequations}\label{eq:regions}
  \begin{align}
\mbox{hard:} &\quad k\sim(1,1,1) \label{eq:khard} \\
\mbox{soft:} &\quad k\sim(\lambda,\lambda,\lambda)\label{eq:ksoft} \\
\mbox{collinear:} &\quad k\sim(\lambda^2,1,\lambda) \label{eq:kcoll}\\
\mbox{ultrasoft:} &\quad k\sim(\lambda^2,\lambda^2,
\lambda^2) \label{eq:kus}
  \end{align}
\end{subequations}
Expanding all master integrals in these regions, the form factors can
be written as a series in $\lambda$ (\ie $z$), in principle to any
order. Details of the calculation of the master integrals in expanded
form are given in Appendix~\ref{sec:integrals}.  As it turns in
Section~\ref{sec:resultstopfac}, restricting this expansion to leading
order in $\lambda$, counting $\log\lambda \sim 1$, the form factors
actually factorize and can be written as a product of hard, soft, and
collinear contributions. All ultrasoft contributions cancel between
diagrams.

\subsection{Dimensional schemes and \texorpdfstring{\neqcd}{neQCD}}

In order to regularize UV and IR divergent integrals, we use
dimensional regularization~({\sc
dreg})~\cite{Bollini:1972ui,tHooft:1972fi} throughout the paper. More
precisely, we formally consider space-time and momentum integrations
in an arbitrary dimension
\begin{align}
 d = 4-2\eps\,
\end{align}
and indicate this shift by a subscript, for example
$\partial^{\mu}_{[4]}\!\to\!\partial^{\mu}_{[d]}$ and
$k^{\mu}_{[4]}\!\to\!k^{\mu}_{[d]}$ for derivatives and loop momenta,
respectively. Depending on the specific realization of \DREG, the
(quasi) dimensionality of other algebraic objects like metric tensors,
$\gamma$ matrices, and vector fields might or might not be different
from $d$. To account for this regularization-scheme dependence, we
write gauge fields associated to particles in 1PI diagrams as
\begin{align}
  A^{\mu,a}_{[4]} \ \to\ A^{\mu,a}_{[d]} + A^{\mu,a}_{[\Neps]}\,,
  \label{eq:Asplit}
  \end{align}
where $A^{\mu,a}_{[d]}$ is a (quasi) $d$-dimensional gauge field and
$A^{\mu,a}_{[\Neps]}$ is a so-called \eps-scalar field. Considering
\Neps as an initially arbitrary quantity, the regularization-scheme
dependence is manifest in terms $\propto\Neps$. In this way it is
possible to consider the most commonly used dimensional schemes
simultaneously, \ie the 't~Hooft-Veltman scheme (\HV)
\cite{tHooft:1972fi}, conventional dimensional regularization (\CDR)
\cite{Collins:1984xc}, the four-dimensional helicity scheme (\FDH)
\cite{Bern:1991aq}, and dimensional reduction (\DRED)
\cite{Siegel:1979wq}. These schemes are defined such that the value of
\Neps is given by 
\begin{subequations}\label{eq:neps}
  \begin{align} 
  \HV, \CDR & :\quad \Neps = 0\,,\phantom{\big|}\\
  \FDH, \DRED & :\quad \Neps = 2\eps\,.\phantom{\big|}
  \end{align}
\end{subequations}
According to the split in \eqref{eq:Asplit}, the QCD covariant
derivative can be written as
\begin{align}
 D^{\mu}_{[4]} \psi_{i}
 \ \to\ \partial^{\mu}_{[d]}\psi_{i}
 + \I \Big( g_s^0 A^{\mu,a}_{[d]} + g_e^0 A^{\mu,a}_{[\Neps]}
 \Big)T^{a}_{ij}\psi_{j}\,,
\end{align}
where $g_e^0$ is the bare coupling of \eps-scalars to fermions. This
so-called evanescent coupling is only present in \FDH\ and \DRED and
has to be introduced at the Lagrangian level as it is not protected by
Lorentz and gauge invariance. Its UV renormalization is therefore
different compared to the one of the gauge coupling $g_s^0$
\cite{Capper:1979ns}. Only \textit{after} UV renormalization, the
numerical values of the renormalized couplings can be set equal. In
the following both couplings are used in the form $\alpha^0_{i}=
(g_i^0)^2/(4\pi)$ with $i\in \{s,e\}$. The corresponding renormalized
couplings are denoted by $\alpha_{i}$.

In what follows we keep the dependence on $\alpha_s$, $\alpha_e$, and
\Neps in the results and refer to this as \neqcd. This allows one to
obtain the corresponding results in all schemes
through~\eqref{eq:neps}.  For a comprehensive definition and a review
of the different dimensional schemes we refer
to~\cite{Gnendiger:2017pys} and references therein.

\section{Results}
\label{sec:resultstop}

In this section we present the result of the form factor \FMms. We begin by
keeping the full mass dependence of $F_1$ and check its IR structure to verify
our calculation. We then discuss the expanded result, which, due to the large
scale separation between $M$ and $m$, is sufficient for most practical
purposes. In fact, we show that the leading term of this expansion represents
an excellent approximation and leads to a factorized form. 

\subsection{Result with the full mass dependence and IR prediction}

We calculated the form factor \FMms at NNLO with full mass dependence
using~\cite{Chen:2018dpt}. The result can be expressed in terms of two
dimensionless variables but involves a huge number of
terms\footnote{The expression is approximately 1GB.}. However, the
master integrals take a particularly simple form if one of the
variables is chosen to be $\chi$, defined as
\begin{align}
\chi+\frac{1}{\chi} = \frac{2\, p\cdot q}{M\, m} = \frac{s}{M\,m}
= \frac{1-x}{z} + z\, .
\label{eq:chichen}
\end{align}
This agrees with $\chi$ as defined in~\cite{Chen:2018dpt}. The
usefulness of $\chi$ can be understood by noting that the
singularities of \FMms can be expressed in terms of UV singularities
of Wilson lines~\cite{Becher:2009kw} with scaling symmetries $p\to
\lambda_1\,p$, $q\to \lambda_2\, q$. Indeed, $\chi$ is invariant under
these symmetries, whereas $s$ is only invariant under the restricted
case $\lambda_1 \lambda_2 = 1$.  More precisely, the poles of \FMms
are governed by the velocity-dependent cusp anomalous dimension, which
is a function of the cusp angle $\beta_{pq} = \log(\chi)$. Therefore,
the poles of \FMms depend only on $\chi$. The finite part, however,
also depends on the second variable $z$.

The result can now be expressed using 540 different GPLs of the form
$G(\{\alpha_i\},1)$, $G(\{\alpha_i\},\chi)$ and $G(\{\alpha_i\},z)$. A
set of eleven letters $\alpha_i$ are needed to express the finite part
of \FMms, namely
\begin{align}
\alpha_i \in \Bigg\{ 
    0, \pm1, \pm z, \pm\frac1z, \pm \Big(z-\sqrt{1-z^2}\Big),
    \frac{1\pm\sqrt{1-z^2}}{z}
\Bigg\}\,.\label{eq:alphabet}
\end{align}

The divergent part contains only simple polylogarithms of $\chi$,
i.e. the alphabet is $\{0,\pm1\}$. In fact, the divergent part can be
checked by considering the IR anomalous dimensions presented
in~\cite{Gnendiger:2014nxa,Gnendiger:2016cpg}. To do this, we
construct $\log({\bf Z})$ as
\begin{align}
\log F_1\Big|_{\text{poles}} = \Bigg\{
                      F_1^{(1)}
                       +  \Bigg[
                        F_1^{(2)} - \frac12 \Big(F_1^{(1)}\Big)^2
                        \Bigg]
                    \Bigg\}_{\text{poles}}+
                    \order{\alpha_i^3}
                    \ \equiv 
                  \  \log({\bf Z}) \,,
\end{align}
where $F_1^{(l)}$ is the $l$-loop contribution to $F_1$.  Using the
explicit expressions for the two-loop heavy-quark anomalous dimension
$\gamma_Q$ and the velocity-dependent cusp anomalous dimension
$\gamma_\text{cusp}(\log\chi)$ in \FDH, as well as the one-loop
$\beta$-functions of \neqcd~(see \cite{Gnendiger:2016cpg} and
references therein), we can predict all poles of $\log({\bf Z})$. The
leading poles expressed in terms of the renormalized couplings of
\neqcd read
\begin{align}
\log({\bf Z}) &= \bigg(\frac{\alpha_s}{4\pi}\bigg)C_F
\frac1{2\eps}\Bigg[
    \frac{\chi^2+1}{\chi^2-1}\log(\chi)-1
\Bigg]\nonumber \\&\quad
+\bigg(\frac{\alpha_s}{4\pi}\bigg)^2\Bigg[
       \frac{1}{\eps^2}\frac{C_FC_A(\Neps-22)+4\,C_Fn_f}{24}\Bigg(
           \frac{\chi^2+1}{\chi^2-1}\log(\chi)-1
\Bigg)+\order{\eps^{-1}}\Bigg]\,.
\label{eq:polesfull}
\end{align}
The result of our explicit computation agrees with this prediction.
Since the full NNLO result of $\log({\bf Z})$ is too complicated to be
listed here it is presented in the ancillary file. The result is given
in \neqcd, but there are no terms involving $\alpha_e$ and no scheme
dependence at the one-loop level.

Beyond the first order in QED with $n_f\!=\!0$, the expression for
$\log \FMms$ is finite since all the divergences of the form factors
are described by exponentiating the one-loop soft poles. Therefore,
after taking the limit $\eps\!\to\!0$, there is no scheme dependence
in $\log\FMms$.

\subsection{Expanded result}

Once the result is expanded in $z\!=\!m/M$, it becomes significantly
simpler. The alphabet~\eqref{eq:alphabet} is now only $\{0,\pm1\}$,
i.e. the result can be expressed completely in terms of the harmonic
polylogarithms~\cite{Remiddi:1999ew}, which are implemented in the
Mathematica package {\tt HPL}~\cite{Maitre:2005uu}. However, some
master integrals can be written in a more compact form by keeping some
specific GPLs with weight two~\cite{Bonciani:2008wf}.

The poles can now be written as
\begin{align}
\log({\bf Z}) &= 
    \bigg(\frac{\alpha_s}{4\pi}\bigg)\frac{C_F}{2\eps} L
    \notag\\[5pt]&\quad+
    \bigg(\frac{\alpha_s}{4\pi}\bigg)^2\Bigg(
        C_AC_F\Bigg\{
            \frac{\Neps-22}{24\eps^2}L
           +\frac1\eps\Bigg[
                \bigg(\frac{67}{36}-\frac{\Neps}{9}-\frac12\zeta_2\bigg)
                        L
                +\frac12-\frac12\zeta_3
            \Bigg]
        \Bigg\}
    \notag\\[5pt]&\qquad\qquad+
        C_Fn_F\Bigg\{
            \frac{1}{6\eps^2}L - \frac{5}{18\eps}L
        \Bigg\}
    \Bigg)+\order{\alpha_i^3,z^2}\,,
\end{align}
    \setlength{\textfloatsep}{28pt}

\begin{figure}[p]
    \centering
    \scalebox{0.7}{
        \centering
        \begin{picture}(430,340)
            \put(20,10){\includegraphics[width=400pt]{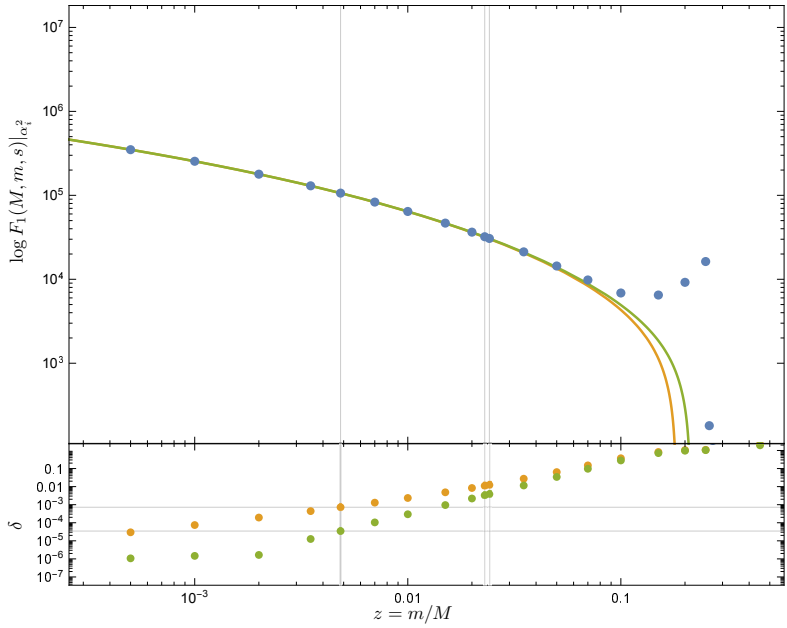}}
            \put(200,2){\small$z=m/M$}
            \put(5,50){\rotatebox{90}{\small$\delta$}}
            \put(5,190){\rotatebox{90}{
                \small$\log\FMms|_{\alpha_i^2}$
            }}
        \end{picture}
    }
    \caption{Comparison of $\log(F_1)|_{\alpha_i^2}$ as a function of
    $z$ in QED at two loops using the exact
    integrals~\cite{Chen:2018dpt} (blue) as well as the expansion at \order{z^0}
    (orange) and \order{z^1} (green) in the upper panel.  The lower
    panel shows the relative difference. The vertical lines indicate
    the physical mass ratios $m_e/m_\mu$, $m_s/m_b$ and $m_b/m_t$.
    }
    \label{fig:numcompz}
\end{figure}
\begin{figure}[p]
        \centering
    \scalebox{0.7}{

        \centering
        \begin{picture}(430,360)
            \put(20,10){\includegraphics[width=400pt]{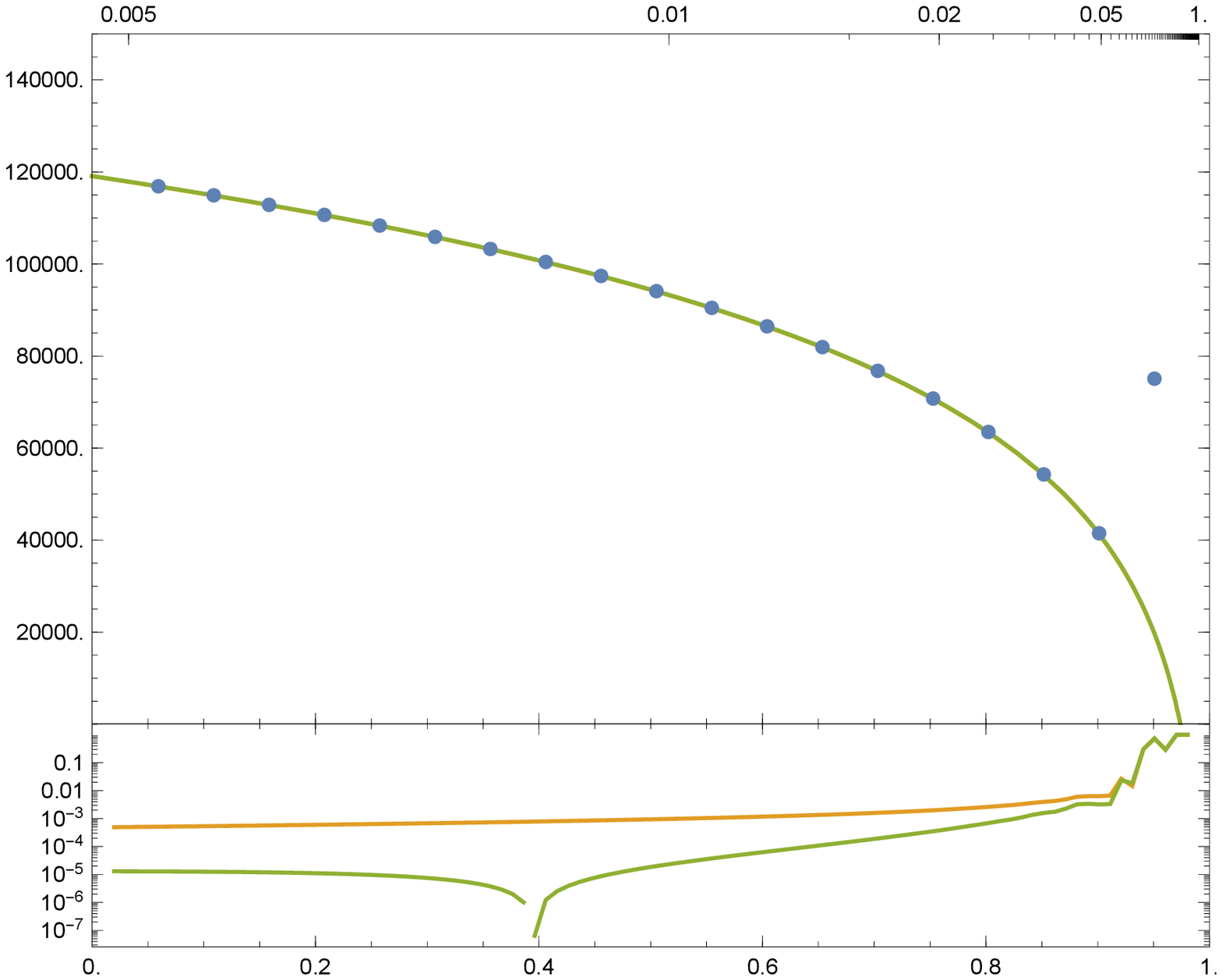}}
            \put(200,2){\small$x=Q^2/M^2$}
            \put(5,50){\rotatebox{90}{\small$\delta$}}
            \put(5,190){\rotatebox{90}{
                \small$\log\FMms|_{\alpha_i^2}$
            }}
            \put(240,340){$\chi$}
        \end{picture}
    }

    \caption{
    The kinematic dependence of $\log(F_1)|_{\alpha_i^2}$ in QED with the full
    mass dependence (blue) as well as the expansion at $\order{z^0}$ (orange)
    and \order{z^1} (green). The lower panel shows the relative difference.
    Note that around $x=0.4$ the deviation between the exact result and the
    \order{z^1} result crosses zero.
    }%
    \label{fig:numcompx}
\end{figure}

\clearpage
\noindent with
\begin{align}
L=\log\Bigg(\frac{s}{m M}\Bigg)-1\,.
\end{align}
The complete expression for \FMms as well as $F_2(M,m,s)$ and $F_3(M,m,s)$ is
provided in an ancillary file attached to this submission.

Using the method of regions, it is not only possible to calculate the leading
terms of \order{z^0} but also subleading terms. In particular, the term of
\order{z^1} can be calculated easily because, apart from trivial terms from the
Dirac algebra, only soft integrals can generate terms odd in $z$.  We note
that, because of the symmetry~\eqref{eq:bermansym}, terms of \order{z^1} drop
out in the full matrix element for decay processes.

We can assess the reliability of our expansion by comparing the full result of
$\log F_1|_{\alpha_i^2}$ to its expansion in QED. We do this in
Figure~\ref{fig:numcompz} numerically, taking into account terms of \order{z^0}
and subsequently of \order{z^1}. The GPLs have been evaluated using an in-house
implementation based on~\cite{Vollinga:2004sn,Frellesvig:2016ske}. At the
physical mass ratio $z\!=\!m_e/m_\mu$, that is particularly relevant for {\sc
MUonE}, the relative error is approximately $10^{-3}$ at \order{z^0} and
$<10^{-4}$ at \order{z^1}. In Figure~\ref{fig:numcompx} we present the
kinematic dependence of \FMms. Note that around $x=0.4$ the difference between
the full mass dependence and the expansion up to \order{z^1} crosses zero. For
values of $x\!\to\! 1$, the agreement gets worse since the expansion in $m^2/s$
breaks down, as can be seen from \eqref{eq:sdef}.

\subsection{Factorization}
\label{sec:resultstopfac}

Motivated by soft-collinear effective theory
(SCET)~\cite{Bauer:2001yt,Bauer:2000yr,Beneke:2002ph}, it is expected that at
leading order in $z\!=\!m/M$ the massive form factor \FMms factorizes into
hard, soft, and collinear parts according to
\begin{align}
\FMms = \sqrt{\Zjet}\times\soft\times \FMzs + \order{m/M}\,.
\label{eq:factorisation}
\end{align}
The hard part corresponds to $\FMzs$ since using the scaling \eqref{eq:khard}
simply results in neglecting $m$ in the integrand.

\begin{figure}[t]
\centering
\scalebox{0.8}{\begin{tikzpicture}
        \centerarc [gluon, line width=0.25mm](0,0)(180:235:1.6)
        \centerarc [gluon, line width=0.25mm](0,0)(270:324:1.6)
        
        \path [draw=black,zigzag it, line
        width=0.25mm](0,0)--(0.8,1.8)node[right]{$Q$};
        
        \draw[line width=.6mm]  (-2.5,0) node[left]{$p$} -- (0,0) ;
        \draw[line width=.6mm,dashed] (0,0)  --(2,-1.5)node[right]{$q$};

        \draw[dashed,line width=.6mm] (-.5,-1.5) circle (0.5cm);
\end{tikzpicture}}
\caption{The $n_m$ bubble giving rise to the soft contribution. Solid
    lines are fermions with mass $M$ and dashed lines are fermions
    with mass $m$.}
\label{fig:diagsoft}
\end{figure}
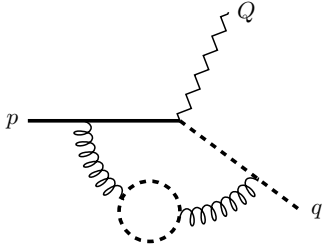

For the soft part $\soft$, we only need to consider diagrams with internal
fermions. Indeed, by performing the formal decoupling of gluon and fermion
fields in the SCET framework, one can show that purely gluonic contributions to
the soft part vanish to all orders~\cite{Bauer:2001yt,Becher:2007cu}. A simple
counting argument implies that only the fermion bubble with mass $m$
contributes~(cf.\ Figure~\ref{fig:diagsoft}). Therefore, the unrenormalized
soft part $\soft^0$ can easily be calculated from first principle in the SCET
framework using \eqref{eq:ksoft}, \ie
\begin{align}
  \label{eq:soft}
\soft^0 = 1+\bigg(\frac{\alpha_s^2}{4\pi}\bigg)\ C_F \int[\D k]\frac{
\big(-2p^\mu\big)\big(-2q_-^\nu\big)
}{(k^2)^2 (2p\cdot k)(2q_-\cdot k)} \Pi_{\mu\nu}^{(n_m)}(k)\,,
\end{align}
where the integral measure is defined in~\eqref{eq:measure}. In accordance with
\eqref{eq:ksoft}, we only use the large component of the collinear momentum
$q$. Because \eps-scalars do not couple to fermions in the eikonal
approximation~\cite{Gnendiger:2016cpg}, there is no contribution
$\propto\Neps$. The function $\Pi_{\mu\nu}^{(n_m)}$ is the contribution of
$n_m$ fermions with mass $m$ to the usual tensorial vacuum polarization. When
calculating $\soft^0$, one encounters an anomaly, \ie the breaking of naive
factorization~\cite{Becher:2010tm,Becher:2011pf}.
Following~\cite{Beneke:2005}, we call this factorization anomaly\footnote{This
is also referred to as collinear anomaly or rapidity
divergence~\cite{Chiu:2011qc}.}. This is a new feature that is only present due
to the large mass $p^2\!=\!M^2$. As we discuss in Section~\ref{sec:resultshq},
if the heavy fermion is replaced by a fermion with mass $m$ or a massless one,
the soft contributions vanish and, therefore, have no factorization anomaly.

The factorization anomaly can be regularized by shifting the power of the
propagator $p\cdot k$ at the diagrammatic level according
to~\cite{Smirnov:1997gx,Becher:2014oda}
\begin{align}
  \label{eq:reganomaly}
\frac1{(k-p)^2-M^2+\io}\to
\frac1{-2p\cdot k+\io} \to (-\nu^2)^\eta\,
\frac1{(-2p\cdot k+\io)^{1+\eta}}\,,
\end{align}
where the regulator $\eta$ has to be expanded before the dimensional regulator
\eps. This regularization also introduces an associated scale $\nu$ that drops
out in the final result. The only further soft contribution is from $n_m$ terms
in the wave-function renormalization of the heavy fermion. Including this
contribution, $Z_2^{\soft}$, we obtain
\begin{align}
\soft = \sqrt{Z_2^{\soft}}\times \soft^0 = 1+\Big(a_s^0(M m)\Big)^2 C_F n_m 
\Bigg[&
  \frac1\eta\bigg(-\frac2{3\eps^2}+\frac{10}{9\eps}
  -\frac{56}{27}-\frac43\zeta_2\bigg)
\notag\\[10pt]&+
\frac1{2\eps^3}
-\frac1{9\eps^2}
+\frac1\eps\bigg(-\frac{26}{27}+\zeta_2\bigg)
\notag\\[10pt]&
+\frac{11}3-\frac23\zeta_3-\frac29\zeta_2
\Bigg]+\order{a_s^3,\eps,\eta}\,,
\end{align}
where we define $a_s^0(x)$ as
\begin{align}
a^0_i(x) = \bigg(\frac{\alpha_i^0}{4\pi}\bigg)
      \bigg(\frac{\mu^2}{m^2}\bigg)^{\eps}
      (-2+\io)^{\eta/2}
      \bigg(\frac{-\nu^2}{x}\bigg)^{\eta/2}\,,
\qquad
i\in\{s,e\}\,,
\end{align}
with an analogous expression for the renormalized couplings.

The calculation of the collinear terms, \Zjet, is more involved because the
integral reduction mixes different momentum regions and is of limited use in
the presence of the regulator~$\eta$.  Therefore, we have explicitly calculated
the collinear gluon contribution to the process by expanding the diagrams using
\eqref{eq:kcoll} without any integral reduction. This way, \Zjet can be written
in terms of the bare couplings of \neqcd as
\begin{align}
\sqrt{\Zjet} &= 1
 +a^0_s C_F \Bigg\{
    \frac1{\eps^2}+\frac1{2\eps}+\zeta_2+2
    +\bigg(4+\frac12\zeta_2\bigg)\eps
    +\bigg(8+2\zeta_2+\frac74\zeta_4\bigg)\eps^2+\order{\eps^3}
\Bigg\}\notag\\&\qquad
 +a^0_e C_F \frac{\Neps}4\Bigg\{
    -\frac1{\eps}-1
    -(1+\zeta_2)\eps
    -(1+\zeta_2)\eps^2+\order{\eps^3}
\Bigg\}\notag\\&\qquad
 +\Big(a^0_s(s)\Big)^2\Bigg\{
    C_F^2\Bigg[
         \frac1{2\eps^4} + \frac1{2\eps^3}
        +\frac1{\eps^2}\bigg(\frac{51}{24}+\zeta_2\bigg)
        +\frac1{\eps  }\bigg(\frac{43}{ 8}-2\zeta_2+6\zeta_3\bigg)
        \notag\\&\qquad\qquad
        +\frac{369}{16}+\frac{61}{4}\zeta_2-18\zeta_4-24\zeta_2\log2-3\zeta_3
    \Bigg]\notag\\&\qquad\quad
   +C_FC_A\Bigg[
        \frac{11}{12\eps^3}   
       +\frac1{\eps^2}\bigg(\frac{25}{9}-\frac12\zeta_2\bigg)
       +\frac1{\eps}\bigg(
            \frac{1957}{216}+\frac{13}2\zeta_2-\frac{15}2\zeta_3
        \bigg)
        \notag\\&\qquad\qquad
       +\frac{31885}{1296}+\frac{38}{3}\zeta_2-13\zeta_4+12\zeta_2\log2
       + \frac{13}3\zeta_3
    \Bigg]\notag\\&\qquad\quad
    -C_FC_A\frac{\Neps}{12}\Bigg[
        \frac1{2\eps^3}+\frac{11}{6\eps^2}
       +\frac1{\eps}\bigg(\frac{215}{36}+3\zeta_2\bigg)
       +\frac{4559}{216}+11\zeta_2+4\zeta_3
    \Bigg]\notag\\&\qquad\quad
    +C_F n_f\frac16\Bigg[
       -\frac1{\eps^3}-\frac{8}{3\eps^2}
       -\frac1{\eps}\bigg(\frac{149}{18}+6\zeta_2\bigg)
       -\frac{3269}{108}-16\zeta_2-8\zeta_3
    \Bigg]\notag\\&\qquad\quad
    +C_F n_m \frac23 \Bigg[
        \frac1\eta\bigg(
            \frac1{\eps^2}-\frac5{3\eps}+\frac{28}9+2\zeta_2
        \bigg)
        -\frac1{\eps^3}+\frac1{2\eps^2}
        +\frac1\eps\bigg(-\frac{55}{24}-3\zeta_2\bigg)
        \notag\\&\qquad\qquad
        +\frac{1675}{432}-2\zeta_2+\zeta_3
    \Bigg]\Bigg\}
\notag\\&\qquad
 +a^0_ea^0_s \Bigg\{
    C_F^2 \frac\Neps4 \Bigg[
        -\frac1{\eps^3}-\frac9{2\eps^2}-\frac{15}{2\eps}
        -23\zeta_2-2\zeta_3+1
    \Bigg]\notag\\&\qquad\quad
    +C_AC_F \frac\Neps8 \Bigg[
        -\frac{11}\eps-\frac{105}2+4\zeta_2+20\zeta_3
    \Bigg]\Bigg\}
\notag\\&\qquad
 +\Big(a^0_e\Big)^2 \Bigg\{
    (C_F\Neps)^2\frac1{32}\Bigg[
        -\frac3{\eps^2}-\frac5\eps-30\zeta_2+\frac{85}2
    \Bigg]\notag\\&\qquad\quad
   +C_F n_f \frac{\Neps}{8} \Bigg[
        \frac1{\eps^2}+\frac7{2\eps}+\frac{21}4+6\zeta_2
    \Bigg]\notag\\&\qquad\quad
   +C_F \Neps \bigg(\frac{C_F}2+\frac{\Neps C_A}{8}-\frac{C_A}4 \bigg)
   \Bigg[
        \frac1{\eps^2}+\frac2{\eps}-3+4\zeta_2
    \Bigg]\notag\\&\qquad\quad
   +C_F \Neps n_m \frac18\Bigg[
        \frac1{\eps^2}+\frac{7}{2\eps}-\frac34+2\zeta_2
    \Bigg]
\Bigg\}+\order{a_i^3,\eps,\eta}\,.
\label{eq:zjet}\end{align} 
The expression for \Zjet is also included in the ancillary file of this
submission. The couplings in~\eqref{eq:zjet} get renormalized in the \MS scheme
with $n_f\!+\!n_m$ flavours. We have confirmed that~\eqref{eq:factorisation}
holds after the heavy $n_h$ flavours are decoupled in
\FMms~\cite{Chetyrkin:1997un,Gnendiger:2016cpg}.

The anomalous $1/\eta$ terms are cancelled between \Zjet and \soft.  However,
because \Zjet contains $s^{-\eta}$, while \soft contains $(m M)^{\eta}$, this
gives rise to new, factorization-breaking logarithms of the form $\log(m M/s)$.
This is related to the breaking of a scaling symmetry between collinear and
soft modes in SCET~\cite{Becher:2010tm,Becher:2011pf}.

Finally, there is the ultrasoft region defined through~\eqref{eq:kus}. This
region contributes to individual master integrals and even diagrams. As
expected, at leading order in $\lambda$, the ultrasoft contributions to the
form factors cancel. Hence, in \eqref{eq:factorisation} no such contribution is
present. 

The factorization \eqref{eq:factorisation} also holds for $F_2$ and $F_3$.
However, these cases are rather trivial at NNLO. Since $F_2$ and $F_3$ vanish
at leading order and the soft contribution only enters at NNLO, there are only
hard and collinear contributions that are non-vanishing. Furthermore, only the
NLO term of \Zjet is required and the factorization anomaly does not enter at
this order. 

\section{Comparison with heavy-quark form factor}
\label{sec:resultshq}

Small mass effects have been considered before in the literature. In
this section we compare our findings to previous results and consider
observables other than the heavy-to-light form factor.

To start with, the collinear contribution, \Zjet, given in
\eqref{eq:zjet} agrees with a corresponding expression obtained
in~\cite{Mitov:2006xs} apart from the $n_m$ terms that were not
considered there. These terms that lead to the factorization anomaly
have been considered in~\cite{Becher:2007cu}, in particular in the
context of the heavy-quark form factor $\Fmms$.
In~\cite{Becher:2007cu} the soft contribution to the heavy-quark form
factor, $\softHQ$, was defined as
\begin{align}
  \softHQ_{\text{\cite{Becher:2007cu}}}  =
  1+\bigg(\frac{\alpha_s}{4\pi}\bigg)^2\ C_F \int[\D k]\frac{
\big(-2p^\mu\big)\big(-2q^\nu\big)
}{(k^2)^2 (2p\cdot k)(2q\cdot k)} \Pi_{\mu\nu}^{(n_m)}(k)\,.
\end{align}
This definition is motivated by the eikonal approximation and does not
lead to a factorization anomaly. Our definition of the soft
contribution to the heavy-quark form factor is motivated by SCET. For
consistency with the collinear contribution, one also has to introduce
the same regulator here. Our definition therefore reads
\begin{align}
  \label{eq:softHQ}
\softHQ &= 1+\bigg(\frac{\alpha_s}{4\pi}\bigg)^2\ C_F \int[\D k]\frac{
    \big(-2p_+^\mu\big)\big(-2q_-^\nu\big)
}{(k^2)^2 (2p_+\cdot k)^{1+\eta}(2q_-\cdot k)} \Pi_{\mu\nu}^{(n_m)}(k)
\,,
\end{align}
where $p$ is assumed to scale anti-collinear and $q$ collinear.
Because any integral of the form
\begin{align}
I(n_1,n_2,n_3) \equiv \int[\D k]\frac{1}{
    (k^2)^{n_1} (2p_+\cdot k)^{n_2} (2q_-\cdot k)^{n_3}}
\Pi_{\mu\nu}^{(n_m)}(k)
\end{align}
depends on $p_+$ and $q_-$ only through $s=2\, p_+\!\cdot q_-$, it is
invariant under simultaneous rescaling $p_+ \to \lambda\ p_+, \ q_-
\to q_-/\lambda$.  This implies $I(n_1,n_2,n_3) =
\lambda^{-n_2}\lambda^{n_3}\ I(n_1,n_2,n_3)$ and, hence, $I=0$ unless
$n_2=n_3$. However, due to the regulator \eqref{eq:reganomaly}, $n_2$
can never be equal to $n_3$. Hence, all occurring integrals vanish and
$\softHQ=1$ at two loops.  For the heavy-quark form factor, the
factorization anomaly in \Zjet is therefore not cancelled by an
anomaly in the soft contribution. In the following we show that,
instead, it is cancelled by an anomaly in \Zantijet. This is a
contribution analogous to \Zjet, but due to the anti-collinear
fermion.

In Section~\ref{sec:resultstopfac} we treat the factorization anomaly
by shifting the propagators of the heavy fermion. Previously, only the
expansion in the collinear direction $e$ contributed because $p^2$ was
large. Now, however, we also have to expand in the anti-collinear
direction~$\bar e$, which results in a new term, \Zantijet.  All
non-anomalous terms in \Zantijet are identical to \Zjet because $e$
and $\bar e$ can be interchanged while simultaneously swapping $p$ and
$q$. This is, however, not valid for the anomalous $n_m$ terms. In
order for the regularization scheme to be consistent, the propagator
that is regularized must remain the same throughout the calculation.
Thereby, the $e\leftrightarrow\bar e$ symmetry is broken for the $n_m$
terms~\cite{Becher:2014oda}. As a consequence, the $n_m$ terms of
\Zantijet have to be obtained through an explicit calculation. We find
\begin{align}
\sqrt{\Zantijet}\Big|_{n_m} = \Big(a^0_s(m^2)\Big)^2 
    C_F n_m \frac23 \Bigg[&
        -\frac1\eta\bigg(
            \frac1{\eps^2}-\frac5{3\eps}+\frac{28}9+2\zeta_2
        \bigg)
        +\frac1{2\eps^3}
        -\frac5{6\eps^2}
        -\frac{253}{72\eps}
        \notag\\&\quad
        +\frac{5083}{432}-\frac{14}{3}\zeta_2-\zeta_3
    \Bigg] + \order{a_s^3,\eps,\eta}
    \,.
\end{align}
This way the relevant expression $\Zjet\times\Zantijet$ is anomaly
free. We can test this result by checking the relation
\begin{align}
  \label{eq:FmmTest}
\Fmms = \sqrt{\Zjet\times\Zantijet}\times\softHQ\times \Fzzs +
\order{m^2/s}\,.
\end{align}
The \CDR result of the heavy-quark form factor \Fmms can, for example,
be obtained from~\cite{Bernreuther:2004ih}, while the massless case
\Fzzs can be found in~\cite{Gehrmann:2005pd}. Further, we have checked
explicitly that \eqref{eq:FmmTest} also holds in \neqcd by using the
\FDH results from~\cite{Gnendiger:2016cpg}.

Because of these differences in the calculation of \softHQ and
\Zantijet, we do not expect agreement beyond the pure QCD
contributions with~\cite{Mitov:2006xs,Becher:2007cu}. However, in
accordance with the previous results, we find
\begin{align*}
Z_{\text{\cite{Mitov:2006xs}}} =
\sqrt{\Zjet\times\Zantijet}\, \Big|_{n_m,\Neps\to0}
\qquad\text{and}\qquad 
Z_{\text{\cite{Becher:2007cu}}} \times \softHQ_{\text{\cite{Becher:2007cu}}} =
    \sqrt{\Zjet\times\Zantijet}\times\softHQ\, \Big|_{\Neps\to0}\,.
\end{align*}

Hence, our results agree with previous ones but extend them to
processes where additional fermions with a large mass are present.
This agreement as well as the fact that \Zjet is the same for \FMms
and \Fmms is a strong indication that the factorization presented here
is general. Thus, it is possible to obtain the leading small-mass
terms of scattering amplitudes such as muon-electron scattering as
follows: The hard part can be obtained from the corresponding
amplitude with $m=0$. For each external fermion of mass~$m$, we
multiply by the corresponding $\sqrt{\Zjet}$ or $\sqrt{\Zantijet}$. In
addition, we have to compute the process-dependent soft contribution
according to the prescription exemplified in~\eqref{eq:soft}
and~\eqref{eq:softHQ}.

\section{Conclusions}
\label{sec:conclusion}

The purpose of this article is to facilitate the future computation of
amplitudes at NNLO for processes with two different non-vanishing
fermion masses. In particular, we are interested in the relation of
such amplitudes for the case where the smaller mass $m$ is kept,
compared to the case where it is set to zero. Since fermion masses in
the Standard Model have a rather strong hierarchy, the full dependence
on $m$ is often not required. This is particularly the case for QED
processes with muons and electrons.

As a starting point, we have calculated the heavy-to-light form
factors at NNLO in QCD (and QED). We have done this by keeping the
full mass dependence as well as by expanding in $m$ by employing the
method of regions. We have verified that the leading term of the
expanded result approximates the full result very well.  For QED we
have also computed the terms of order $m$, resulting in an even better
approximation.

At leading order in this expansion the relation between the massless
and massive results can be written in a factorized form. The presence
of a second large mass leads to a factorization (or collinear) anomaly
in the soft contribution. Hence, our scheme for the soft part differs
from previous expressions. We have also independently calculated the
mass factorization functions \Zjet and \Zantijet from the
heavy-to-light form factor and the heavy-quark form factor,
respectively. This has been done by identifying the collinear
contribution to the form factors at the diagrammatic level and
evaluating them. Our expressions for \Zjet and \Zantijet agree with
previously obtained results~\cite{Mitov:2006xs, Becher:2007cu} for the
purely gluonic terms, where no anomaly is present. Combining our
expressions of soft and (anti-)collinear contributions, the
factorization anomaly cancels in all cases. We have also reproduced
previously discussed factorized expressions for leading small-mass
terms in the case where only one fermion mass is present. Taken
together, this provides strong evidence that our factorization is
universally valid for processes with two different non-vanishing
masses with a strong hierarchy.

A particular application we have in mind is to obtain the leading
electron mass terms of the NNLO amplitude for muon-electron
scattering. This is relevant for the proposed {\sc MUonE} experiment.
The additional ingredients required are the NNLO amplitude for
massless electrons, that corresponds to the hard part, as well as the
process-specific soft contribution. We have provided a consistent
definition of the latter contribution and found that its computation
is much simpler than the calculation of the hard part.

\acknowledgments{ We are grateful to Sophia Borowka, Seraina Glaus,
  and Michael Spira for fruitful discussions on numerical evaluation
  of loop integrals. We want to further thank Andrea Visconti for
  sharing his notes on the \FDH calculation of \FMzs. TE and YU are
  supported by the Swiss National Science Foundation (SNF) under
  contract 200021\_178967 and 200021\_163466, respectively.

}

\appendix

\section{Master integrals}
\label{sec:integrals}

The determination of the form factors~\eqref{eq:ff} requires the
calculation of 40 scalar two-loop master integrals that are 
presented in this appendix. The integrals are graphically defined in
Figure~\ref{fig:MasterIntegrals} and normalized to
\begin{align}
[\D k] = \Bigg(\frac{\mu^2}{M^2}\Bigg)^{\eps}
        \Gamma(1-\eps) \frac{\D^dk}{\I \pi^{d/2}}\,.
\label{eq:measure}
\end{align}
One integral, {\tt 6TopP1}, is not fully defined this way because it
contains a numerator. It is explicitly given by
\begin{equation}\begin{split}
\begin{gathered}
\begin{tikzpicture}[scale=.7,baseline={(0,0)},rotate=90]
		\path [draw=black,zigzag it,line width=0.25mm] (0,.52) -- (0,0);
		\draw [line width=0.6mm] (-1,-1.732)--(0,0);
		\draw [line width=0.6mm, dashed] (1,-1.732)--(0,0);
		\draw [snake it, line width=0.25mm] (0.85,-1.472)--(-0.425,-0.736);
		\draw [snake it, line width=0.25mm] (-0.85,-1.472)--(0.425,-0.736);
		\node at (0,.8) {\scriptsize $\mathcal{N}$};
\end{tikzpicture}
\end{gathered}&= \int [\D k_1][\D k_2]
\frac{(k_1+k_2)^2}{
    [k_1^2            ]
    [ k_2^2           ]
    [(k_1-p)^2-M^2    ]
    [(k_1+k_2-p)^2-M^2]
}\\&\qquad\qquad\times\frac{1}{
    [(k_2-q)^2-m^2    ]
    [(k_1+k_2-q)^2-m^2]
}\, .
\end{split}\label{eq:6topp1}\end{equation}

\newcommand{\inputmaster}[2]{
    \begin{minipage}[b][2.43cm][b]{0.165\textwidth}
        \centering
        \scalebox{0.89}{\input{figures/#2}}\\
        {\scriptsize \texttt{#1}}\\[5cm]
    \end{minipage}
}

\begin{figure}[p]
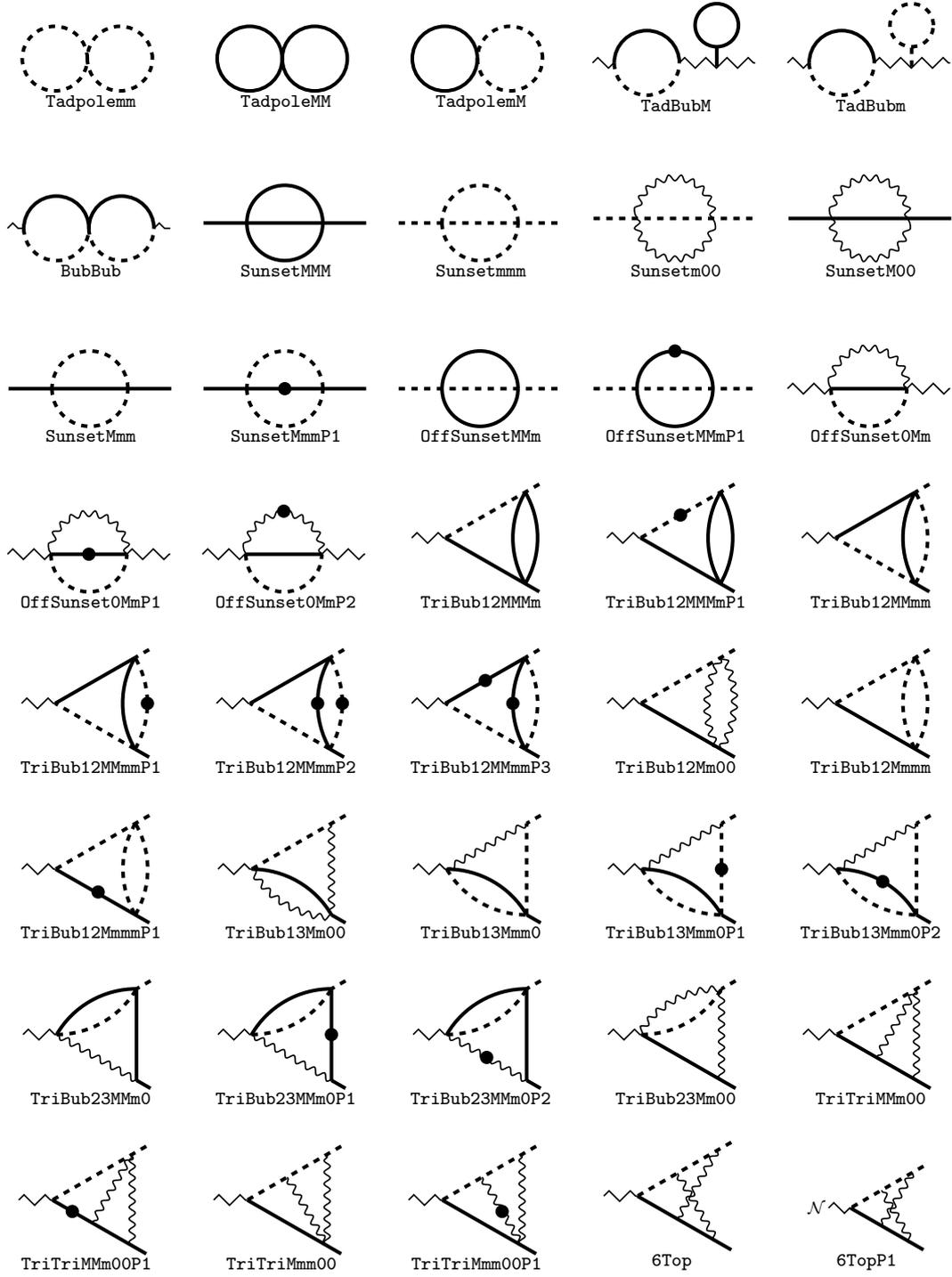

   \centering
\inputmaster{Tadpolemm}{masterTadpolezz}
\inputmaster{TadpoleMM}{masterTadpoleMM}
\inputmaster{TadpolemM}{masterTadpolezM}
\inputmaster{TadBubM}{masterTadBubM}
\inputmaster{TadBubm}{masterTadBubz}
\inputmaster{BubBub}{masterBubBub}
\inputmaster{SunsetMMM}{masterSunsetMMM}
\inputmaster{Sunsetmmm}{masterSunsetzzz}
\inputmaster{Sunsetm00}{masterSunsetz00}
\inputmaster{SunsetM00}{masterSunsetM00}
\inputmaster{SunsetMmm}{masterSunsetMzz}
\inputmaster{SunsetMmmP1}{masterSunsetMzzP1}
\inputmaster{OffSunsetMMm}{masterOffSunsetMMz}
\inputmaster{OffSunsetMMmP1}{masterOffSunsetMMzP1}
\inputmaster{OffSunset0Mm}{masterOffSunset0Mz}
\inputmaster{OffSunset0MmP1}{masterOffSunset0MzP1}
\inputmaster{OffSunset0MmP2}{masterOffSunset0MzP2}
\inputmaster{TriBub12MMMm}{masterTriBub12MMMz}
\inputmaster{TriBub12MMMmP1}{masterTriBub12MMMzP1}
\inputmaster{TriBub12MMmm}{masterTriBub12MMzz}
\inputmaster{TriBub12MMmmP1}{masterTriBub12MMzzP1}
\inputmaster{TriBub12MMmmP2}{masterTriBub12MMzzP2}
\inputmaster{TriBub12MMmmP3}{masterTriBub12MMzzP3}
\inputmaster{TriBub12Mm00}{masterTriBub12Mz00}
\inputmaster{TriBub12Mmmm}{masterTriBub12Mzzz}
\inputmaster{TriBub12MmmmP1}{masterTriBub12MzzzP1}
\inputmaster{TriBub13Mm00}{masterTriBub13Mz00}
\inputmaster{TriBub13Mmm0}{masterTriBub13Mzz0}
\inputmaster{TriBub13Mmm0P1}{masterTriBub13Mzz0P1}
\inputmaster{TriBub13Mmm0P2}{masterTriBub13Mzz0P2}
\inputmaster{TriBub23MMm0}{masterTriBub23MMz0}
\inputmaster{TriBub23MMm0P1}{masterTriBub23MMz0P1}
\inputmaster{TriBub23MMm0P2}{masterTriBub23MMz0P2}
\inputmaster{TriBub23Mm00}{masterTriBub23Mz00}
\inputmaster{TriTriMMm00}{masterTriTriMMz0}
\inputmaster{TriTriMMm00P1}{masterTriTriMMz0P1}
\inputmaster{TriTriMmm00}{masterTriTriMzz0}
\inputmaster{TriTriMmm00P1}{masterTriTriMzz0P1}
\inputmaster{6Top}{master6Top}
\inputmaster{6TopP1}{master6TopP1}

   \caption{
    The 40 scalar two-loop master integrals that contribute to the
    heavy-to-light form factors. For the definition of the numerator of
    \texttt{6TopP1}, see~\eqref{eq:6topp1}. The solid and dashed thick
    lines correspond to the heavy and light fermions, respectively.
    The wavy and zigzag lines correspond to massless gauge bosons and the
    external current. The heavy (light) fermion momentum $p$ ($q$) is
    taken to be incoming (outgoing). A dot indicates that the
    corresponding propagator is squared.
    }
    
   \label{fig:MasterIntegrals}
\end{figure}

We calculated these integrals using the method of
regions~\cite{Beneke:1997zp}. Motivated by
SCET~\cite{Bauer:2000yr,Bauer:2001yt,Beneke:2002ph}, we expect hard,
soft, and collinear regions to appear. In fact, individual integrals
can contain ultrasoft regions that drop out of the final result at
leading order in $z$. The
existence of such regions has been noted before~\cite{Smirnov:1999bza}
as has the fact that they drop out of the final
result~\cite{Becher:2007cu}.

There are two equivalent formulations of the method of regions:
expanding the momentum representation of the integral in momentum
regions using light-cone coordinates or expanding in the alpha representation
after Feynman parametrization~\cite{Smirnov:1999bza}.  In the former,
the $k_+$, $k_-$, and $k_\perp$ components of the loop momentum are
scaled differently and then expanded. This way one obtains the regions
defined in~\eqref{eq:regions}.  In the alpha representation one
instead scales the Feynman parameters.  The revelation of regions in
this representation has been automatized in the Mathematica
program \texttt{asy}~\cite{Pak:2010pt,Jantzen:2012mw} based on
geometric properties of scaleless integrals. We used both methods
concurrently.

The leading hard contribution of integrals always corresponds to the
matching massless integral~\cite{Bell:2006tz,Ferroglia:2013dga}.
Subleading terms can also be matched using a Passarino-Veltman
decomposition and integral reduction.

Collinear, soft, and ultrasoft contributions can be obtained using
Mellin-Barnes decomposition. In our case, the techniques developed
in~\cite{Smirnov:1999gc,Tausk:1999vh} were sufficient to calculate all
integrals. For more complicated (multiple) Mellin-Barnes integrals, we
used the programs {\tt MB.m}~\cite{Czakon:2005rk}, {\tt
MBresolve.m}~\cite{Smirnov:2009up}, and {\tt
barnesroutines.m}~\cite{Kosower:2007} to resolve and simplify the
integrals and, in rare cases, sum residues with {\tt
XSummer}~\cite{Moch:2001zr,Moch:2005uc}. When encountering
hypergeometric functions, we expanded them using {\tt
HypExp}~\cite{Huber:2005yg} to harmonic
polylogarithms~\cite{Remiddi:1999ew,Maitre:2005uu}.

All expanded integral solutions were numerically compared with
the full integrals~\cite{Chen:2018dpt} as well as the numerical
program {\tt FIESTA}~\cite{Smirnov:2015mct} for various values of $z$.

The analytic expression expanded in $m$ for all 40 integrals can be
found in the ancillary file. 

\clearpage

\bibliographystyle{JHEP}
\bibliography{muon_ref}{}

\end{document}